\documentclass[11pt]{article}
\usepackage{bbm}
\usepackage{amsmath,amsthm,amssymb}
\usepackage{graphicx}
\usepackage{tikz}
\usepackage[utf8]{inputenc}
\usepackage{pgfplots}

\usepackage{setspace}
\usepackage{pict2e,color}
\usepackage{multirow}
\usepackage{booktabs}
\usepackage{amsfonts}
\usepackage{verbatim} %for adding comments
\usepackage{sectsty}
\usepackage{url}
\usepackage{subfigure}
\usepackage{empheq}
\usepackage[normalem]{ulem}
\usepackage[titletoc,title]{appendix}
\usepackage[flushleft]{threeparttable}
\usepackage{soul} % use this (many fancier options)

\usepackage[numbers]{natbib}
 \bibpunct[, ]{(}{)}{,}{a}{}{,}%
 \def\BIBand{and}%
 
\usepackage{algorithm,algorithmic}

\usepackage{lscape}
\usepackage{algorithmic}
\usepackage{algorithm}
\usepackage{graphicx}
\usepackage{graphics}
\usepackage{threeparttable}
\usepackage{bm}
\usepackage{subfigure}
\usepackage{booktabs}
\usepackage{xcolor}
\usepackage{multicol}
\usepackage{multirow}
\usepackage{verbatim}

\usepackage{amsmath}
\usepackage{etoolbox}
\preto\subequations{\ifhmode\unskip\fi}
\usepackage{booktabs}
\usepackage{multirow}

\theoremstyle{definition}

\title{\Large An Optimization-and-Simulation Framework for Redesigning University Campus Bus System with Social Distancing}

\begin{document}
%\graphicspath{{figures/}}

\allowdisplaybreaks
\author{Gongyu Chen, Xinyu Fei, Huiwen Jia, Xian Yu\thanks{Department of Industrial and Operations Engineering, University of Michigan at Ann Arbor, USA. Email: {\tt chgongyu,xinyuf,hwjia,yuxian@umich.edu};}
    ~~~Siqian Shen\thanks{Corresponding author; Department of Industrial and Operations Engineering, University of Michigan at Ann Arbor, USA. Email: {\tt siqian@umich.edu}.}}
\date{}

\maketitle

\begin{abstract}
The outbreak of coronavirus disease 2019 (COVID-19) has led to significant challenges for schools, workplaces and communities to return to operations during the pandemic, requiring policymakers to balance individuals' safety and operational efficiency. In this paper, we present our work using mixed-integer programming and simulation for redesigning routes and bus schedules for University of Michigan (UM)'s campus bus system during the COVID-19 pandemic. We propose a hub-and-spoke design and utilize real data of student activities to identify hub locations and bus stops to be used in the new routes. Using the same total number of buses to operate, each new bus route has 50\% or fewer seats being used and takes maximumly 15 minutes, to reduce disease transmission through expiratory aerosol. We sample a variety of scenarios that cover variations of peak demand, social-distancing requirements, and break-down buses, to demonstrate the system resiliency of the new routes and schedules via simulation. The new bus routes are implemented and used by all UM campuses during the academic year 2020-2021, to ensure social distancing and short travel time. Our approach can be generalized to redesign public transit systems with social distancing requirement during the pandemic to reduce passengers' infection risk. 
\end{abstract}
   


\section{Introduction}
\label{sec:intro}

The University of Michigan (UM)’s fleet of buses provides an estimated 8 million rides between its three campuses in 2019 \citep{bluebus-news1}. 
In an effort to design a safe campus bus system for the Fall semester of 2020 in light of COVID-19, researchers at UM simulated how aerosol particles exhaled from passengers sitting in any seat would travel through the interior of a campus bus  \citep[see][]{bluebus-aerosols}. 
In that work, physics-based computer models of aerosol dispersion were developed under different social-distancing and bus-operating conditions, and the results were validated with experiments using water vapor in a real bus. To minimize passengers' possible exposure to the virus, all bus passengers and drivers must wear face coverings, and the capacity of buses also needs to be reduced to half of the original capacity or even lower, while the length of each trip needs to be limited to 15 minutes or less. 

The pre-COVID campus bus system at UM uses long routes designed to minimize transfers. Some routes (e.g., the original Commute South, Commute North and Northeastern Shuttle) take 45 minutes to 1 hour to travel even without including the passenger loading/unloading time at each bus stop. Simply breaking down the original long routes into shorter ones will significantly increase the number of buses needed to cover all the original demand, especially when only half bus capacity can be utilized. Meanwhile, it is not possible to add more buses and drivers due to the high cost of purchasing a campus bus (approximately \$800,000 each) and the difficulty of retaining and hiring qualified drivers during the pandemic, according to our conversation with the UM campus transit operational team \citep{bluebus-chat}. The goal here is to design a safe UM bus transit system with all routes being shorter than 15 minutes, to satisfy student, faculty and staff on-campus travel demands, using the same number of buses but having 50\% or less capacity.

To accomplish the goal, we develop a mathematical optimization model and a simulation platform to design a hub-and-spoke system that relies on short, direct routes with fewer stops. The new routes not only limit all the rides to 15 minutes or less, but also enable more frequent service using the same number of buses, and therefore reduce passenger wait time as well as the number of passengers on each bus. We validate the results via simulation by generating a diverse set of scenarios that take into account spatial-temporal variations of demand, changes of social-distancing requirements, and random bus break-downs (or equivalently driver no-shows). The simulation results not only show that we achieve the goal given by UM bus operators, but also verify that all the riders can find a bus stop within 5 minutes of walking distance and 70\% of passengers still take only one bus transfer for completing their trips. For 35\% of passengers, the average waiting time is less than 5 minutes. For 70\% of passengers, the average waiting time is less than 10 minutes. Through simulation, we also identify the routes that have higher utilization on average or during peak hours, and the bus stops that have longer average waiting time. To improve the system resilience, we develop contingency plans of moving buses to these routes from others in case of some bus failures. The new bus routes are used by UM during the academic year 2020-2021 \citep{current-bus-route, bluebus-news2}, when the COVID vaccines were still being developed or not available to all. Students were able to spend less time on buses with fewer riders, as compared to if the pre-COVID routes were in use.

\section{Literature Review}
\label{subsec:literature}

Our work is related to prior work on transit network planning, including  
transit network design \citep[see, e.g.,][]{bb-13-yu2012transit,bb-5-ouyang2014continuum}, 
%Frequency Setting (FS) \citet[see, e.g.,][]{},
transit network timetabling \citep[see, e.g.,][]{bb-3-ibarra2012synchronization,bb-6-fonseca2018matheuristic}, 
vehicle scheduling problem \citep[see, e.g.,][]{bb-12-naumann2011stochastic,bb-14-yan2012robust},
and transit network evaluation \citep[see, e.g.,][]{bb-26-chandrasekar2002simulation,bb-18-zhang2019data}.

In particular, the transit network design (TND) problem determines the lines, types of vehicles, and stop spacing to meet population's movement requirements. \citet{bb-13-yu2012transit} use a direct traveler density model \citep[first proposed in][]{yang2007parallel} to maximize demand density of route under resource constraints. \citet{bb-4-kim2014integration} propose a probabilistic analytical model to design and synchronize a bus transit network by integrating conventional and flexible bus services and coordinating bus arrivals at transfer terminals. 
We refer to \citet{bb-24-guihaire2008transit} for a comprehensive review of TND problems, where some limiting the lengths of routes in the network design, similar to the 15-minute travel limit in our problem.

In terms of transit network planning (TNP), \citet{bb-10-desaulniers2007public} conduct a survey on mathematical methods used for individual steps of strategic, tactical, operational planning, as well as real-time control related to TNP. 
\citet{bb-25-ibarra2015planning} comprehensively review the TNP literature and real-time control strategies suitable to bus systems. \citet{ceder2016public} summarizes efficient solutions of TNP, discuss how to model operations in practice including bus loading/unloading time and passenger behavior, based on the author's rich experience of research and working with practitioners in this field. 

The  methods we use for redesigning campus bus routes are also closely related to hub-spoke system design \citep[also referred to as Hub Location Problems, see, e.g.,][]{campbell2012twenty}. 
The early work on hub-spoke system design is initiated by \citet{o1986location, o1986activity}, and hub-spoke system design is also referred to as the hub location problem. 
\citet{campbell2002hub} and \citet{farahani2013hub} provide comprehensive review on mathematical models, solution methods, and applications of hub-spoke systems.
\citet{campbell2012twenty} reflect on the origins of hub location research, especially in transportation, provide some commentary on the present status of the field, and discuss some promising directions for future effort.

\section{Mathematical Models for Route Design}
\label{sec:route}

\subsection{Data Collection}
 
We formulate a mathematical programming model to configure new UM bus routes based on a hub-and-spoke design. To to do this, we first gather all the data in the 2019-2020 school year from the department of student services, including student activities and utilization of gyms, dinning halls, libraries, etc., locations of student dorms, the main bodies of students who live in on-campus and off-campus residencies, their majors, school years, and typical course schedules for different colleges and schools at UM. Because some staff and faculty also use the campus buses to travel from their parking lots to their departments, we gather the information about parking lots' locations and their capacities from the department of transportation and parking at UM.  
We also gather all the UM medical school's related data including their faculty, staff, and students' course schedules, time of shifts in the hospital, to estimate the demand from the medical school. Using all the data, we estimate the average demand between pair-wise locations on campus during different hours of each school day (from Monday to Friday). We associate these locations to existing bus stops in the old UM bus system (which are 110 stops), and identify all the bus stops that one can use within 5-minute walking distance to each location.

For this section, we identify three hub locations respectively in the three UM campuses, which have the largest origin or destination travel demand during most hours everyday. In later sections, the same real-data sources will be used to estimate demand between origins and destinations in all UM campuses, for determining bus frequency and evaluating performance of each route using simulation.

\begin{figure}[ht!]
    \centering
    \includegraphics[width=0.8\textwidth]{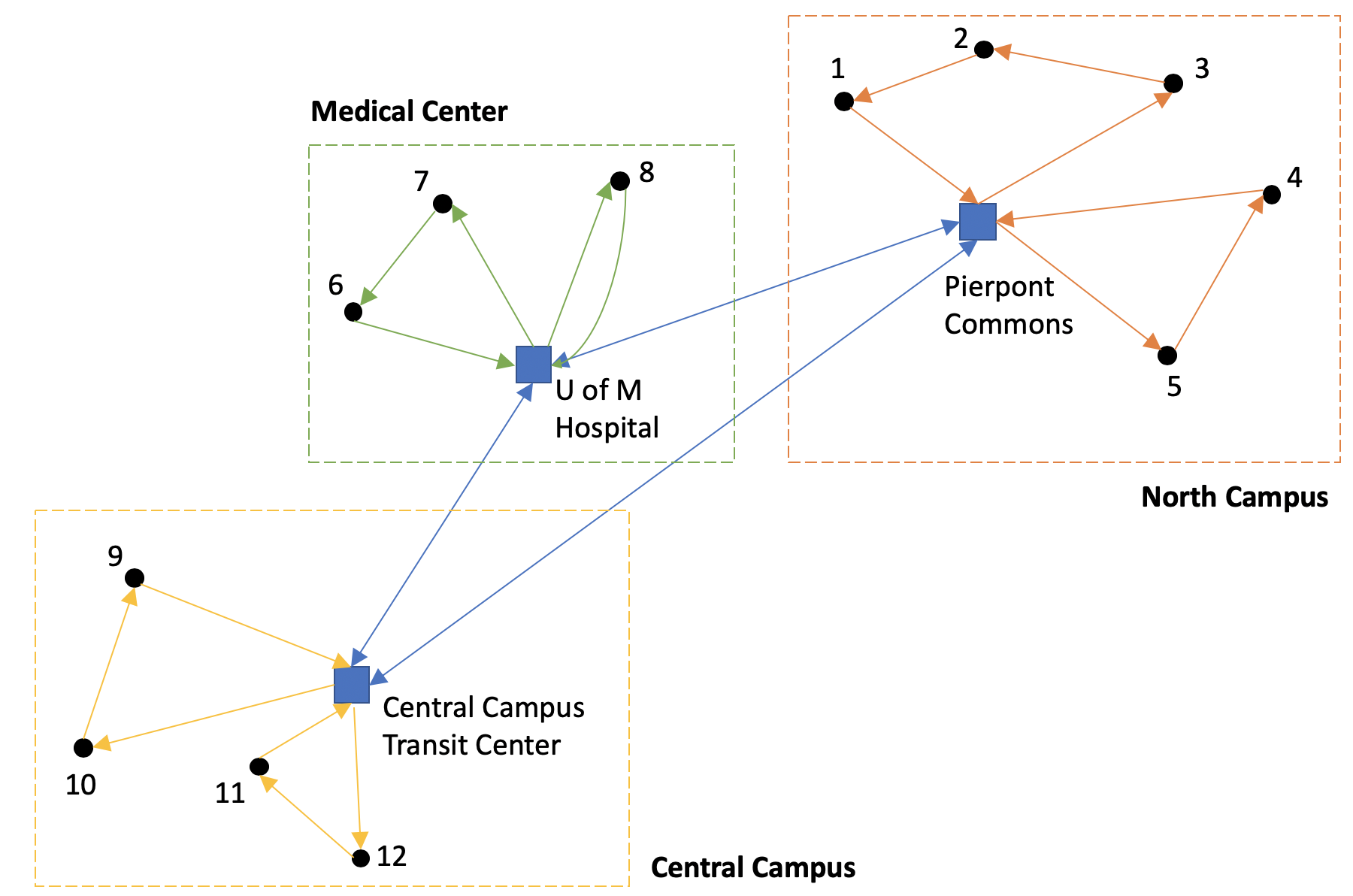}
    \caption{An illustration of the hub-and-spoke system for operating UM buses in all its Ann Arbor campuses
    }
    \label{fig:hub}
\end{figure}
In Figure \ref{fig:hub}, we show an abstract hub-and-spoke system with three hub locations (i.e., the squares) in the UM North Campus, Central-South Campus and Medical Campus, respectively, having the highest travel demand during peak hours in 2019-2020. In our new design, a Campus Connector route will connect these hubs with very frequent buses. Next, we design shorter routes to connect other popular bus stops within each of the three campuses (i.e., sub-regions) from and to each hub. 

\subsection{Mixed-Integer Programming Model}
We define notation and formulate an integer programming model for designing bus routes originating from and returning to each hub location as follows. Let $I$ and $J$ be the sets of potential bus-stop locations and routes that start and end in one of the hubs, respectively. For each route in $J$, we denote $K$ as the sequence of locations visited on the route. Our model aims to assign a bus-stop location from $I$ (or a hub location) to each visit in $K$ for each route that is in use. Note that the number of bus stops on each route is not necessarily the same, and $|K|$ is set as the maximum number of visits allowed in any route.

Let $c_j$ be the cost of operating route $j$ for all $j\in J$. According to our collaborator (i.e., UM campus transit system operators), they normally hire a certain number of bus drivers (including back-up drivers) for each route created, and drivers' salary and benefits are the main part in the fixed cost $c_j$. Other parts in $c_j$ include bus-stop setup and bus maintenance cost. 
We denote $t_{i_1,i_2}$ as the average travel time from stop $i_1$ to stop $i_2$ for all $i_1,\ i_2\in I\cup\{\textrm{hub}\}$, $\tilde{t}_{\text{stop}}$ as the average loading/unloading time at each stop, and obtain their values from our collaborator. We assume that both $t_{i_1,i_2}$ and $\tilde{t}_{\text{stop}}$ are deterministic and known, validated by our collaborator based on their experience of operating campus buses in our university town. The $\tilde{t}_{\text{stop}}$ is assumed to be a constant for each stop, rather than a function of the number of riders we load/unload, following the advice from our collaborator.

We define binary variables $x_j,\ y_{ij},\ u_{i,k}^j,\ u_{\textrm{hub},k}^j\in\left\{0,1\right\}$ for all $i\in I,\ j\in J,\ k\in K$ such that $x_j=1$ if we operate route $j$, $y_{ij}=1$ if we assign bus stop $i$ to route $j$, $u_{i,k}^j=1$ if bus stop $i$ is the $k^\textrm{th}$ visit in route $j$, and $u_{\textrm{hub},k}^j=1$ if the hub is the $k^\textrm{th}$ visited location in route $j$, respectively.

Using the defined parameters and variables, we first present the overall objective function of the optimization model as: \footnotesize
\begin{equation}
\min \sum_{j\in J} c_j x_j + \alpha 
    \sum_{j\in J}\left(\sum_{i\in I}t_{\textrm{hub},i}u_{i,1}^j+\sum_{k=1}^{|K|-1}\sum_{i_1,i_2\in I\cup\left\{\textrm{hub}\right\}}t_{i_1,i_2}u_{i_1,k}^ju_{i_2,k+1}^j+\sum_{i\in I}t_{i,\textrm{hub}}u_{i,|K|}^j + \tilde{t}_{\text{stop}}\left(|K| - \sum_{k=1}^{|K|}u_{\textrm{hub},k}^j\right) \right), \label{eq:obj-0}
\end{equation}
\normalsize
where coefficient $\alpha$ is used as a weight to convert the total travel time and time spent on all stops, into a related  cost of operations and its value can be chosen based on gas price, operational frequency and other factors that will affect the operational cost, as compared to the overall fixed cost $c_j$ in the first term. Note that the objective function uses a similar form in the facility location problem where there is a fixed cost associated with whether or not to operate a route and a continuous cost associated with the travel time of all routes. We use parameter $\alpha$ to adjust the emphasis on the two types of costs.

The objective function \eqref{eq:obj-0} involves bilinear terms $u_{i_1,k}^j u_{i_2,k+1}^j$, which we define by new variables $z_{i_1,i_2,k}^j$ and $z_{i_1,i_2,k}^j=u_{i_1,k}^j u_{i_2,k+1}^j$ can be linearized using McCormick envelopes \citep{mccormick1976computability}:  
\begin{subequations}
%\label{mc-eq}
\begin{align}
\label{mc-eq-1}
    & z_{i_1,i_2,k}^j\le u_{i_1,k}^j,\ \forall i_1,i_2\in I\cup\left\{\textrm{hub}\right\}, \ k=1,\ldots, |K|-1,\ j\in J,\\
\label{mc-eq-2} & z_{i_1,i_2,k}^j\le u_{i_2,k+1}^j,\ \forall i_1,i_2\in I\cup\left\{\textrm{hub}\right\},\ k=1,\ldots, |K|-1,\ j\in J,\\
\label{mc-eq-3} & z_{i_1,i_2,k}^j\ge u_{i_1,k}^j+u_{i_2,k+1}^j-1,\ \forall i_1,i_2\in I\cup\left\{\textrm{hub}\right\},\ k=1,\ldots, |K|-1,\ j\in J.
\end{align}
\end{subequations}

For each sub-region that contains one hub, the mixed-integer programming model for designing new routes is given by: 
\begin{subequations}
\footnotesize
\label{ts-mip}
\begin{align}
\min\quad&\sum_{j\in J} c_jx_j + \alpha 
    \sum_{j\in J}\left(\sum_{i\in I}t_{\textrm{hub},i}u_{i,1}^j+\sum_{k=1}^{|K|-1}\sum_{i_1,i_2\in I\cup\left\{\textrm{hub}\right\}}t_{i_1,i_2}z_{i_1,i_2,k}^j +\sum_{i\in I}t_{i,\textrm{hub}}u_{i,|K|}^j + \tilde{t}_{\text{stop}}\left(|K| - \sum_{k=1}^{|K|}u_{\textrm{hub},k}^j\right) \right) \label{eq:obj}\\
\text{s.t.}\quad
& y_{ij}\le x_j,\ \forall i\in I,\ j\in J,\label{eq:c3}\\
& \sum_{k\in K}u_{i,k}^j=y_{ij},\ \forall i\in I,\ j\in J,\label{eq:c4}\\
& \sum_{i\in I}u_{i,k+1}^j \le \sum_{i\in I}u_{i,k}^j, \ \forall k=1,\ldots, |K|-1,\ j\in J,\label{eq:c6}\\
& \sum_{i\in I}u_{i,k}^j+ u_{\textrm{hub},k}^j= x_j,\ \forall k\in K,\ j\in J,\label{eq:c5}\\
& \sum_{i\in I}t_{\textrm{hub},i}u_{i,1}^j+\sum_{k=1}^{|K|-1}\sum_{i_1,i_2\in I\cup\left\{\textrm{hub}\right\}}t_{i_1,i_2}z_{i_1,i_2,k}^j+\sum_{i\in I}t_{i,\textrm{hub}}u_{i,K}^j + \tilde{t}_{\text{stop}}(|K| - \sum_{k=1}^{|K|}u_{\textrm{hub},k}^j)\le \mathcal{T}, \ \forall j\in J,\label{eq:time}\\
& \mbox{\eqref{mc-eq-1}--\eqref{mc-eq-3},} \nonumber\\
& x_j\in\left\{0,1\right\},\ y_{ij}\in\left\{0,1\right\},\ u_{i,k}^j, \ z_{i_1,i_2,k}^j\in\left\{0,1\right\},\ \forall i\in I,\ k\in K,\ j\in J. \label{eq:c9}
\end{align}
\end{subequations}
\normalsize 
The objective function \eqref{eq:obj} linearizes the original objective \eqref{eq:obj-0}. Constraints \eqref{eq:c3} allow bus stop $i$ being assigned to route $j$ if route $j$ is in use; Constraints \eqref{eq:c4} indicate that if a bus stop is in use (i.e., it is assigned to some route by having $y_{ij} = 1$), then it can be only assigned to one visit in the route. (Note that we do not necessarily need to assign a bus stop to a route and a bus stop can also belong to multiple routes.) 
Constraints \eqref{eq:c6} prohibit assigning a bus stop to a visit in a route if an earlier visit has not been filled by any bus stop.  Constraints \eqref{eq:c5} indicate that for any visit in a route that does not have any bus stop being assigned to it (i.e., $\sum_{i\in I}u_{i,k}^j = 0$ at some $k$), the hub location is assigned to the visit (i.e., $u_{\textrm{hub},k}^j = 1$). (Note that together with Constraints \eqref{eq:c6}, if $u_{\textrm{hub},k^*}^j = 1$ starting from some visit $k^*$ in a route $j$, then Constraints \eqref{eq:c5} will enforce $u_{\textrm{hub},k}^j = 1$ for all visits $k > k^*$ on the same route. That is, the bus will be staying in the hub for the remaining visits $k^*, k^*+1,\ldots,|K|$.) Following the result of Constraints \eqref{eq:c6} and \eqref{eq:c5}, we can use $|K| - \sum_{k=1}^{|K|}u_{\textrm{hub},k}^j$ to compute the actual number of bus stops in any route $j$. Therefore, the left-hand side of Constraints \eqref{eq:time} adds the total travel time and stop time on each route and the constraints enforce a time limit $\mathcal{T}$ (e.g., 15 minutes) for finishing each route.

\subsection{Route Solution Description} 
We code Model \eqref{ts-mip} using Python and use the Gurobi solver 9.0.3 for optimizing the routes for each UM campus. Our numerical tests are conducted on a Windows 2012 Server with 128 GB RAM and an Intel 2.2 GHz processor.  In total, we are given 110 bus stops from the pre-COVID bus system and Model \eqref{ts-mip} will determine the bus stops we will use in the routes in North and Central campuses and how to connect them. (Note that one of the hubs in the UM medical campus are connected by the Campus Connector between the two hubs in North and Central campuses, and its nearby bus stops are assigned to routes associated with the other two hubs.) 

\begin{figure}[ht!]
    \centering
    \includegraphics[width=0.5\textwidth]{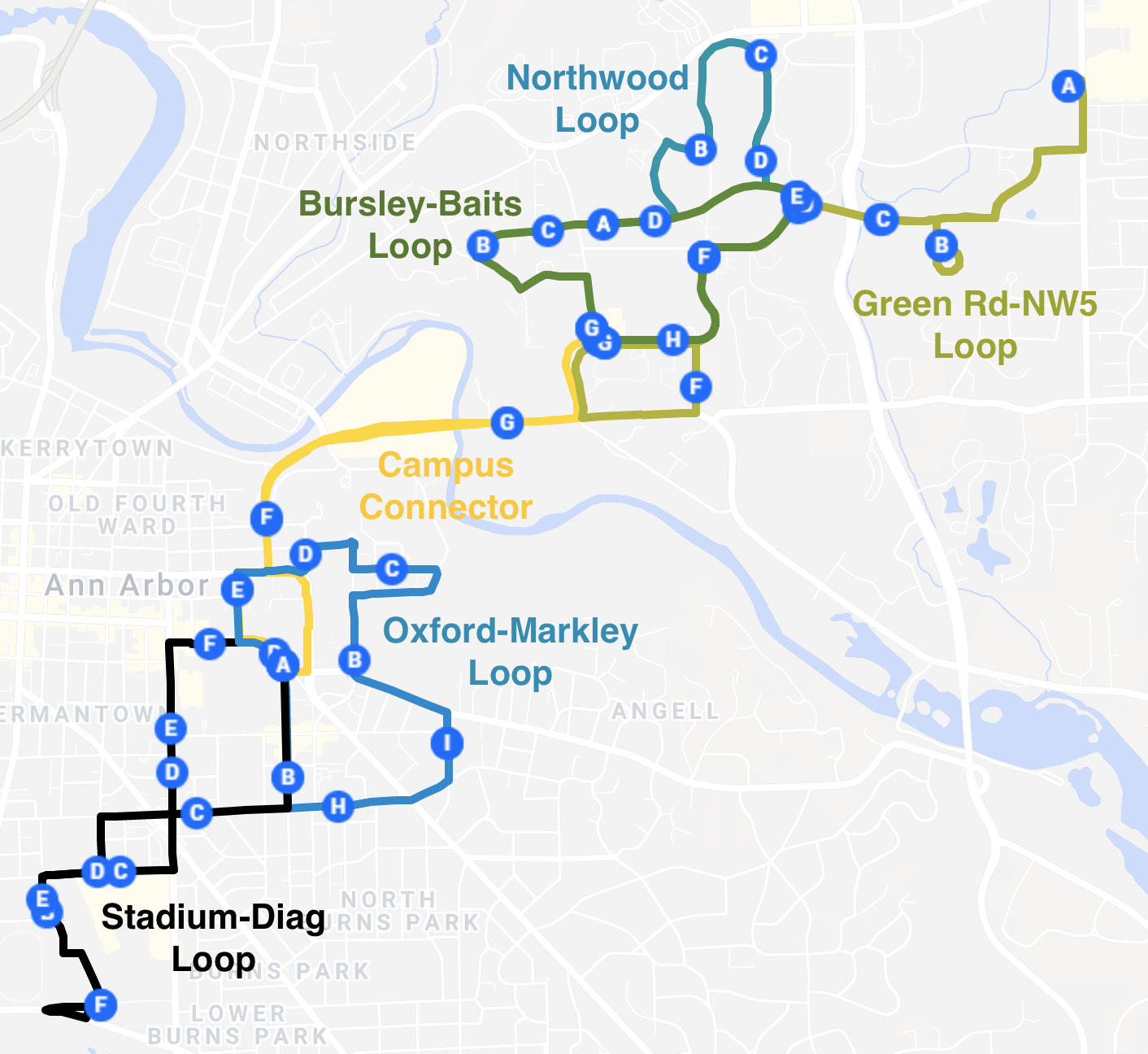}
    \caption{Optimal routes obtained by solving Model \eqref{ts-mip} for designing a hub-and-spoke system.}
    \label{fig:combined}
\end{figure}

\begin{figure}[ht!]
	\centering
	\subfigure[Campus Connector]{%
		\includegraphics[width=0.31\textwidth]{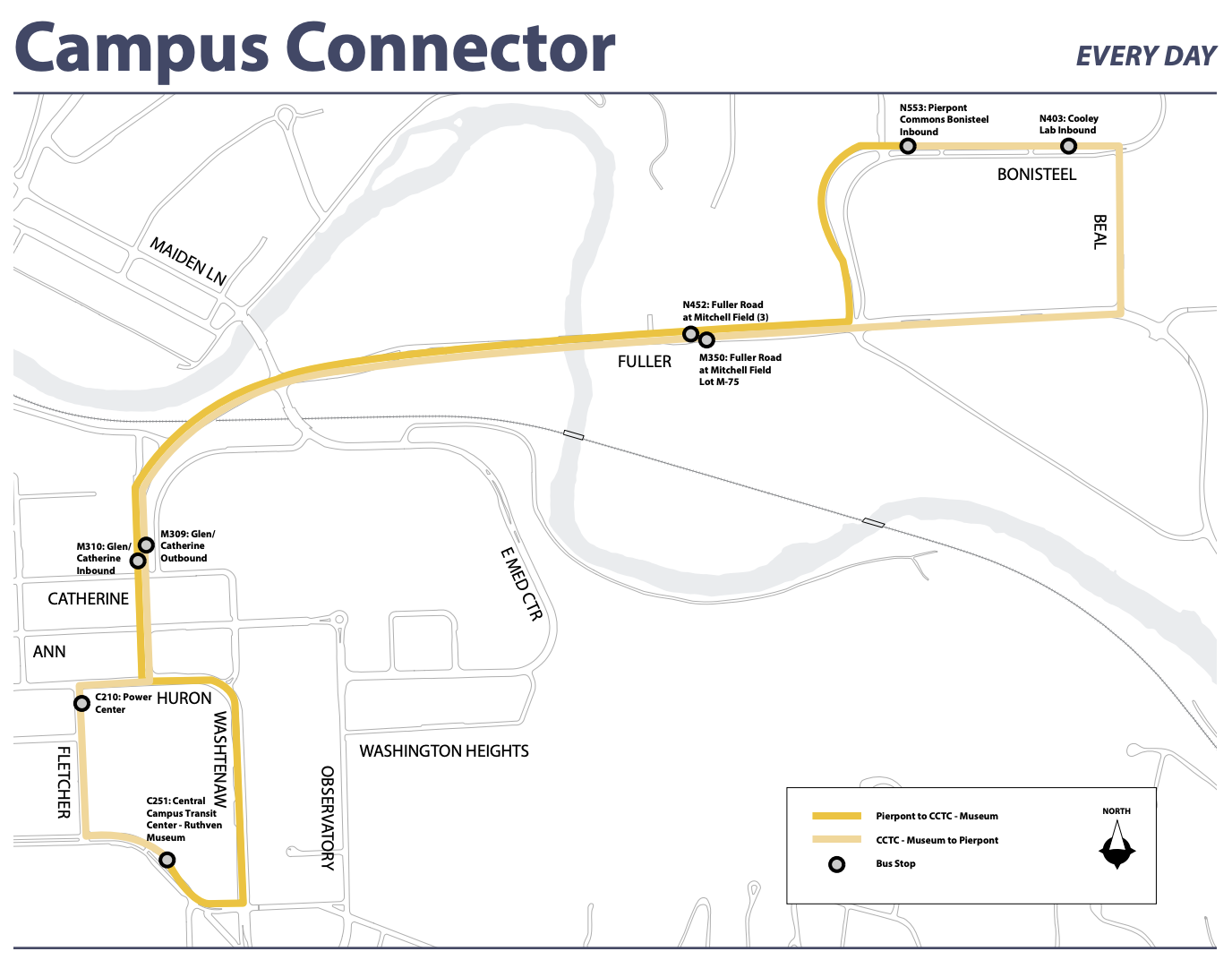}
	}
	\subfigure[Stadium-Diag Loop]{%
	\includegraphics[width=0.31\textwidth]{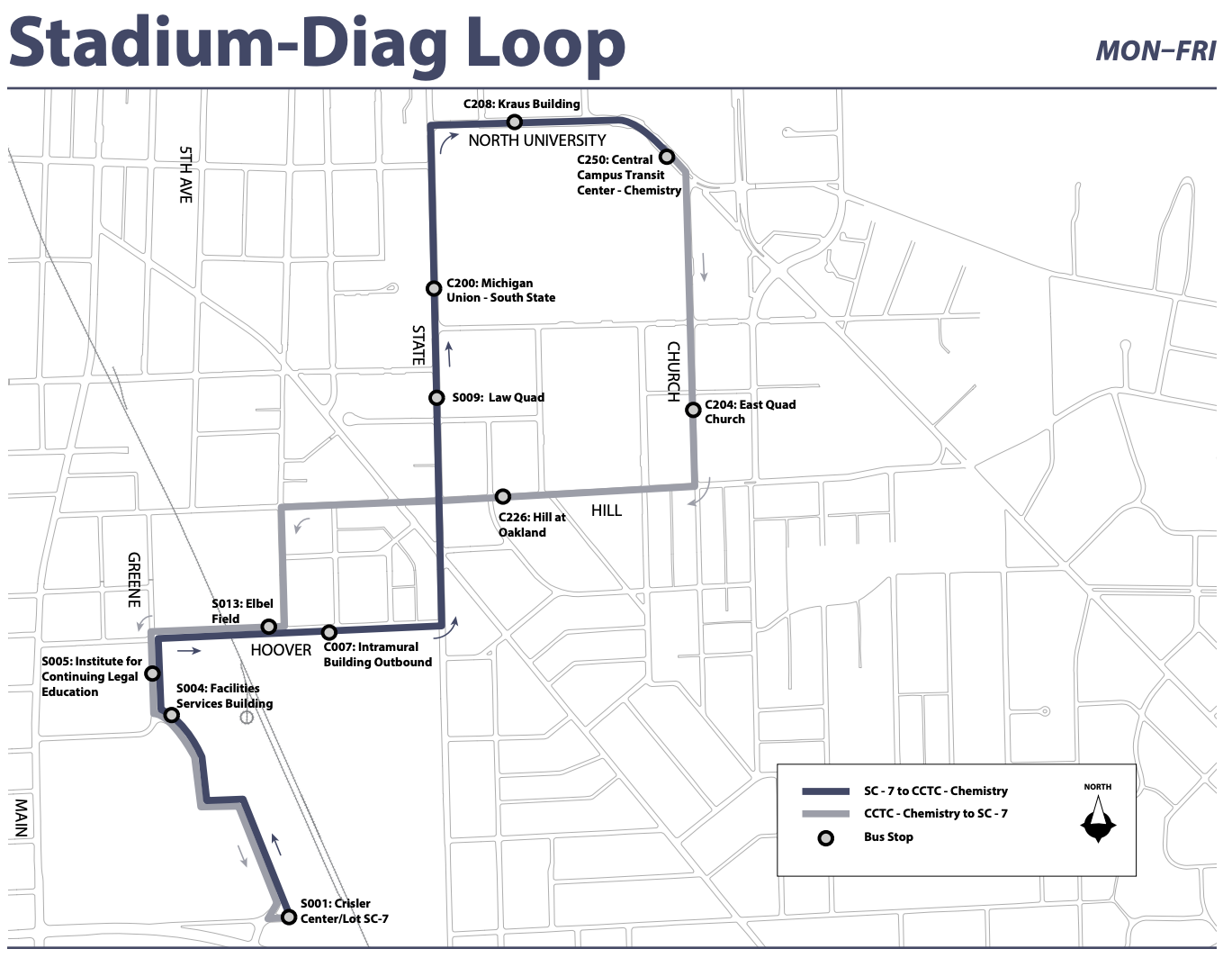}
	}
	\subfigure[Oxford-Markley Loop]{%
	\includegraphics[width=0.31\textwidth]{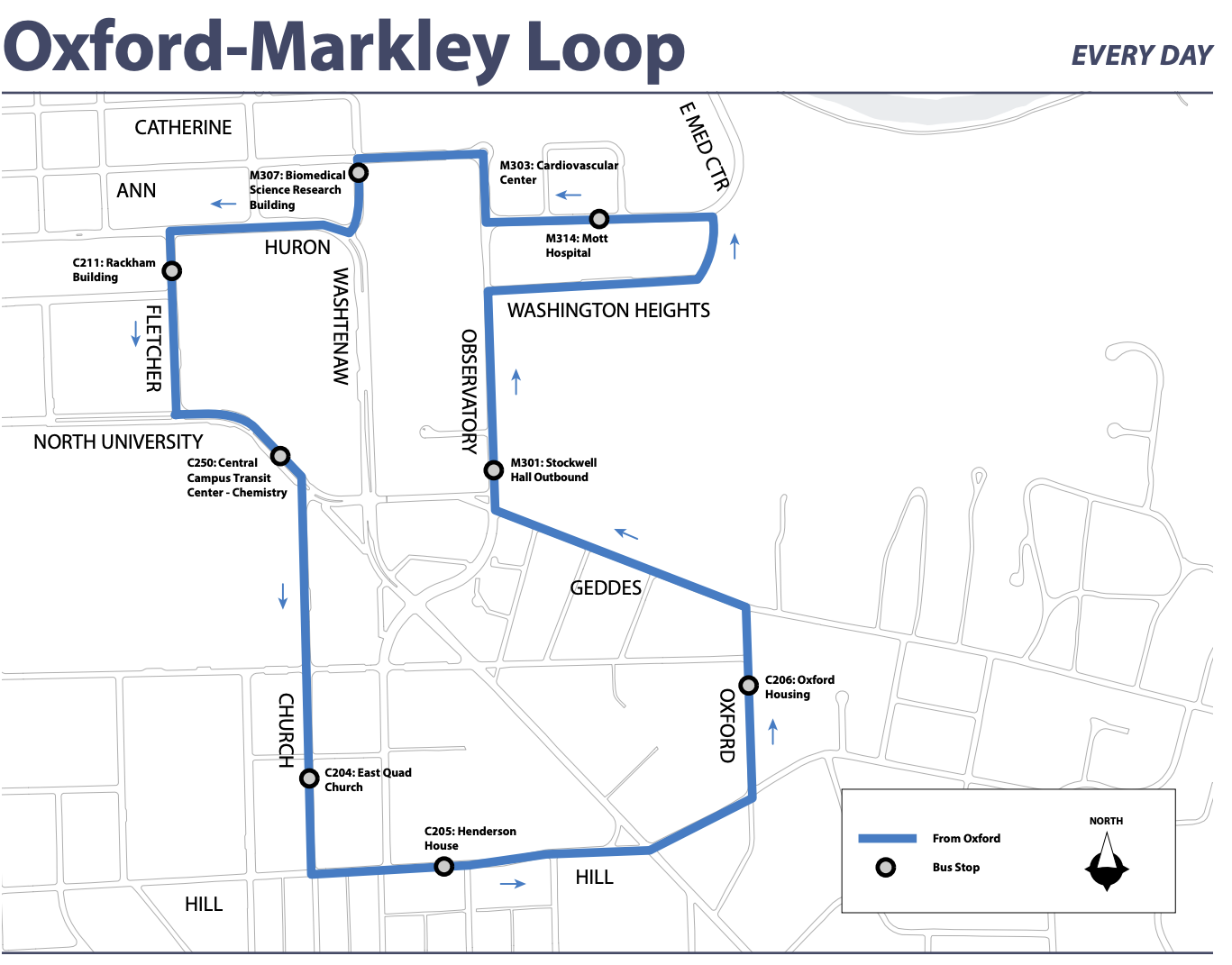}
	}
	\subfigure[Bursley-Baits Loop]{%
	\includegraphics[width=0.31\textwidth]{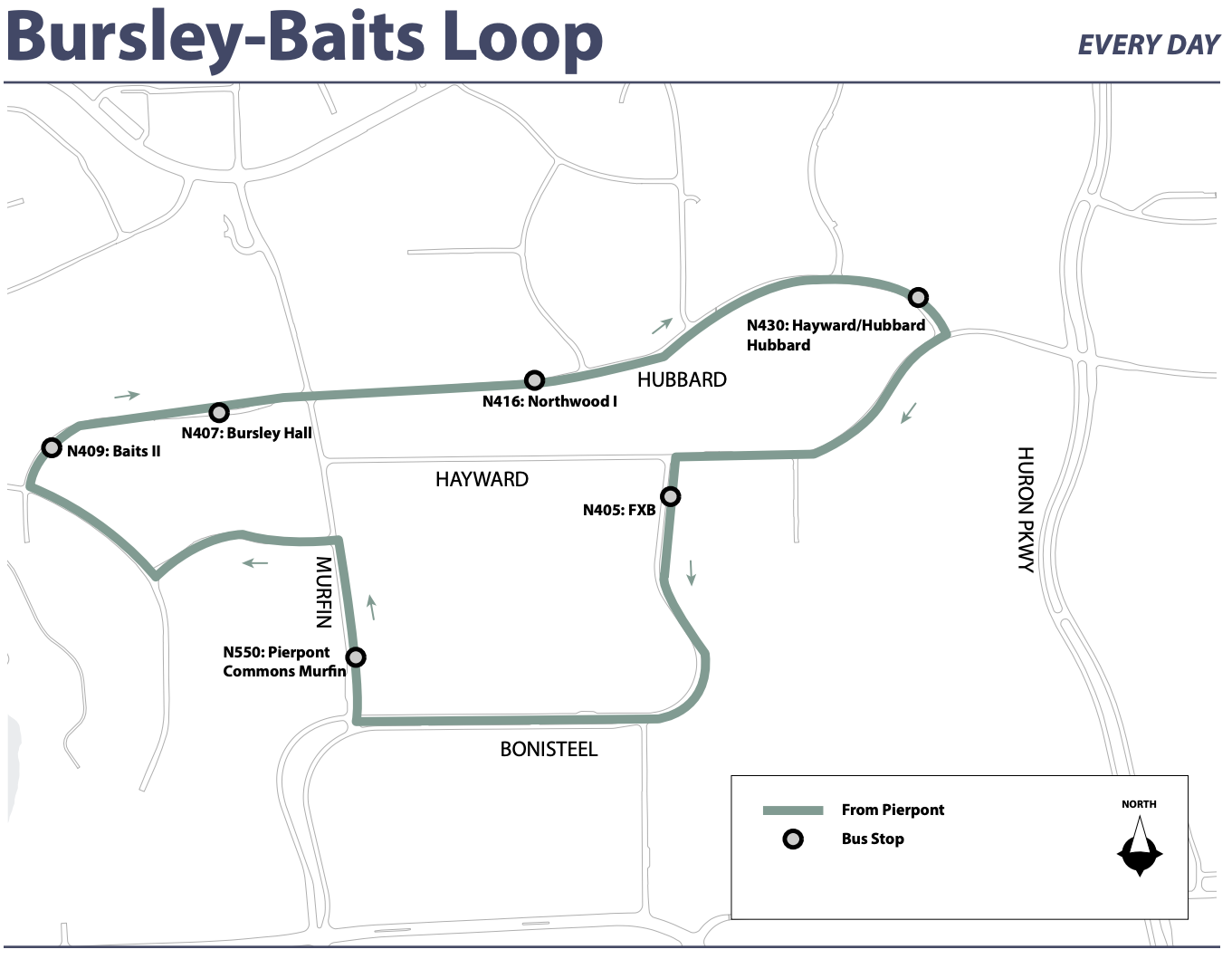}
	}
		\subfigure[Northwood Loop]{%
	\includegraphics[width=0.31\textwidth]{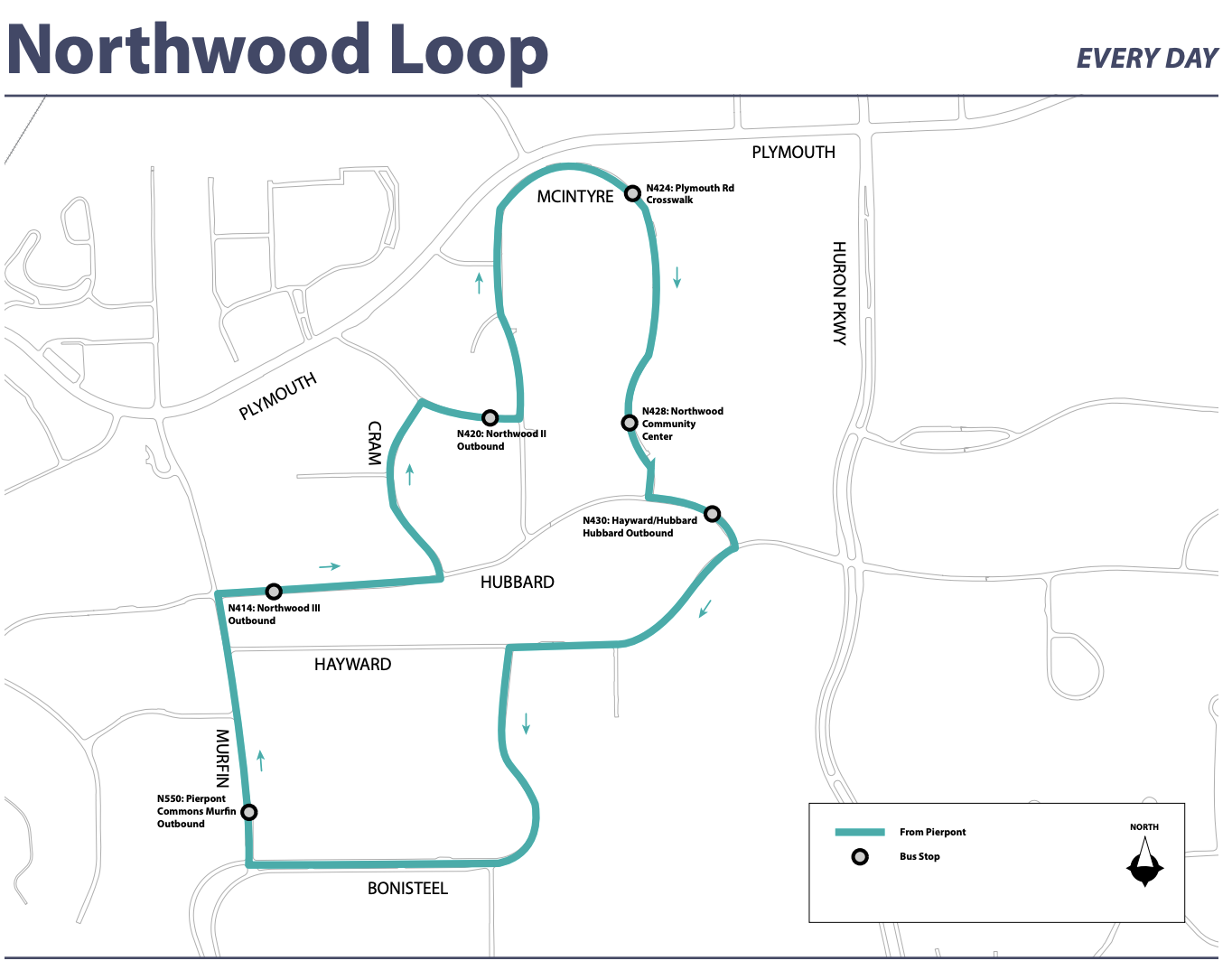}
}
	\subfigure[Green Rd-NW5 Loop]{%
	\includegraphics[width=0.31\textwidth]{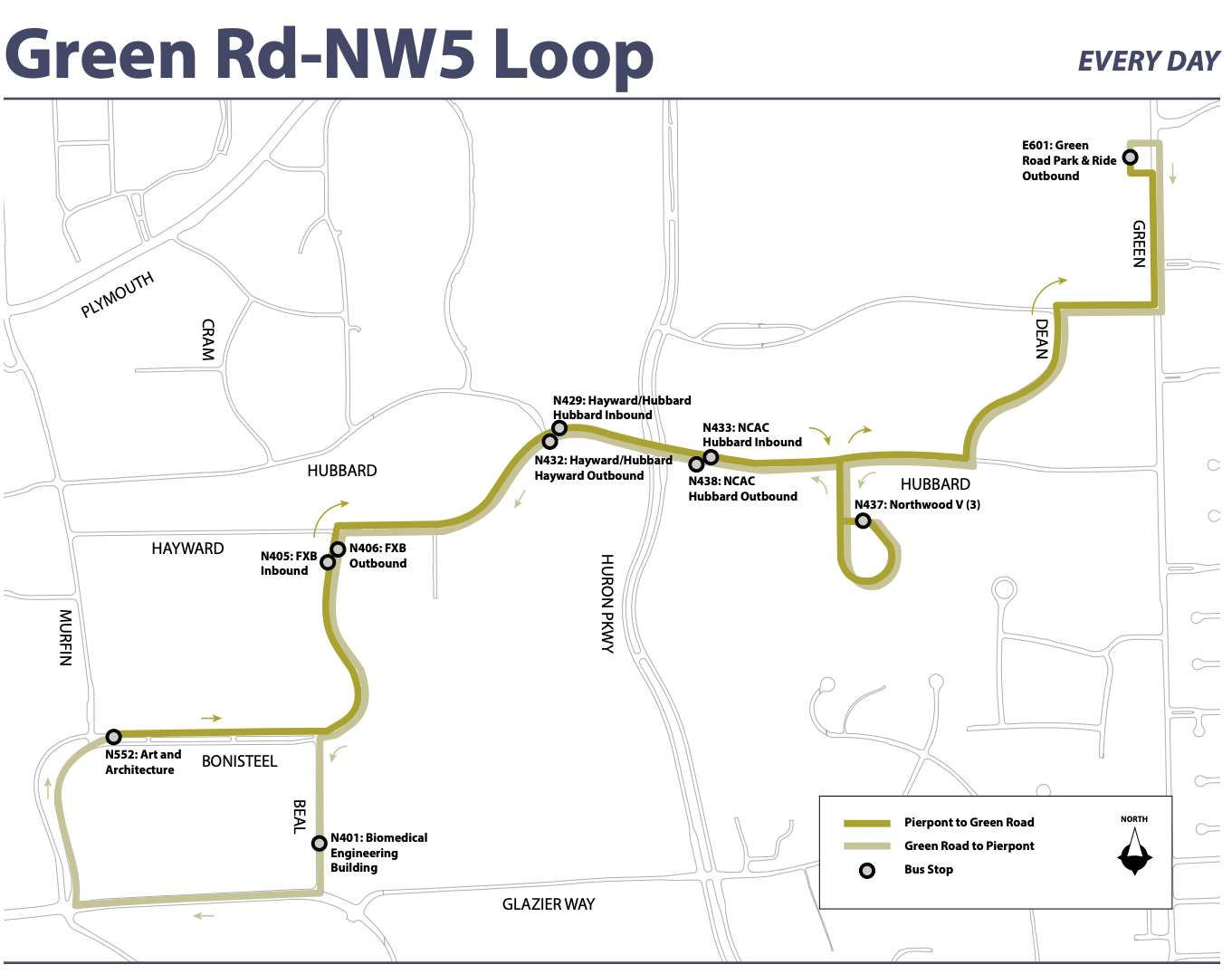}
	}
\caption{Individual routes and bus stops used in the UM 2020-2021 bus transit system.}
\label{fig:individual}
\end{figure}

After solving Model \eqref{ts-mip}, there are in total 50 stops picked to form five different routes (three in North campus and two in Central campus), together with the route that links all the hubs. 
We present optimal routes produced by Model \eqref{ts-mip} in Figure \ref{fig:combined} and individual routes with detailed stops in Figure \ref{fig:individual} for operating UM campus buses during the pandemic. Among these six routes, Bursley-Baits Loop, Northwood Loop and Green Rd-NW5 Loop operate in the north campus of UM, while Oxford-Markley Loop and Stadium-Diag Loop operate in the central campuses. The Campus Connector operates in between respective hubs in the north and central campuses. The CPU times for finding three routes in North campus and two routes in Central campus are 11.19 seconds and 9.74 seconds, respectively. 

\subsection{Bus Frequency Design}
%\label{sec:frequency}

Next, we describe how to design the number of buses and frequency needed for each new route shown in Figures \ref{fig:combined} and \ref{fig:individual}. The goal is not to increase the number of needed buses or drivers, but to meet the same demand levels as our old bus routes can before COVID. We are given two types of buses, one is a shuttle with 35 seats that was used on the old North-East route pre-COVID, and the other is a bus with 70 seats that was used on other routes in the old bus system. The bus stops on our new Green Rd-NW5 Loop overlap with the ones on the old North-East route and therefore, we will use the 35-seat shuttles only on the Green Rd-NW5 Loop. 

We consider 50\% of the total capacity of either a shuttle or a bus can be used in the new system, to meet the social distancing requirement. For each new route, we first obtain its estimated single-trip traveling time (as the sum of the total driving time and passenger loading/unloading), and then examine all the old bus routes that pass through each stop in the new route, to calculate an approximate frequency provided by buses in the old system that the new route needs to match. Note that not all the bus stops in each new route have the same frequency in the old system, and we use the one that has the highest frequency to match. For example, suppose that a bus stop has two old routes covering it, each running every 20 minutes. Then during one hour, the old system can take $70\times 6=420$ passengers from that bus stop at maximum because there is one bus arriving at the stop every 10 minutes. If the stop is only involved in one bus route in the new system, because we can only use 50\% of the total bus seats, we need $12$ buses running every hour to match the $420$-per-hour riding capability and the new route should run every $60/12 = 5$ minutes. 

For some bus routes, the bus frequency varies depending on different times of a day in the old system, and we perform the above analysis for all bus stops and their corresponding new routes during all peak hours. Some bus stops may be involved in more than one route in the new system, and we make sure that the added capability of all passing new routes can match the capability of the old routes at these stops. Because the small number of bus routes involved in either the old or the new system, we are able to easily and quickly design the frequency of buses on each new route following the above procedures. 

Finally, we verify that the numbers of buses (with 70 seats) and shuttles (with 35 seats) needed for fulfilling the desired frequency on each new route do not exceed the total number of buses and shuttles in the current system. We also make sure that the number of available drivers during different shifts between 6:30am and 1:30am (the next day) can cover all the buses needed in those shifts. The driver scheduling is completed by a third-party company that the UM transit department hires and is not in the scope of the problem we consider in this paper. If more drivers are desired for certain periods, the University usually hires student drivers at the beginning of each semester as backup drivers. 

\section{Result Validation and Improvement Using Simulation}
\label{sec:simulation}

\subsection{Framework and Main Logic}

We validate and improve the performance of the new transit system using discrete-event simulation. 
For each bus stop, we define three related areas: wait area (WA), load area (LA), and unload area (UA). Buses arrive at each hub according to given frequency during different hours and move cyclically along the route paths. At each stop, buses unload passengers at UA, move to LA and load passengers from WA, and then leave for UA of the next stop along the route path. A passenger entity, who either gets off a bus or newly enters the system, will arrive at UA of the stop. The entity will decide whether to leave the system, to wait at WA, or to move to a nearby stop for taking another bus (i.e., a transfer) for continuing a trip. A passenger moves from WA to LA when a bus with sufficient capacity arrives and if the passenger needs to get on the bus. We summarize ``Variables'' and entity-specified ``Attributes'' of the simulation model in Table~\ref{attr}, which will be used in our descriptions of the simulation framework in Algorithms \ref{alg_main} and \ref{alg_func}. 
\begin{table}[htbp]
  \centering
  \caption{Attributes and variables of the simulation model}
    \resizebox{0.7\textwidth}{!}{%
    \begin{tabular}{l}
    \hline
    \textbf{Variables} \\
    \hline
    stopID: ID associated with each in-use bus stop;\\
    currStop: stopID of the bus stop where an entity currently stays;\\
    maxCap: the maximum number of passengers a bus can hold.\\
    \hline
    \hline
    \textbf{Bus-related Attributes}\\
    \hline
    currCap: number of passengers in a bus of interest;\\
    Route: route ID of the bus.\\
    \hline
    \hline
    \textbf{Passenger-related Attributes} \\
    \hline
    origin: stopID of the origin bus stop of a passenger;\\
    dest: stopID of the destination stop that the passenger is moving towards.\\
    \hline
    \end{tabular}
    }
  \label{attr}
\end{table}

The logic process described above is detailed in Algorithm~\ref{alg_main}. Passengers enter the bus system by arriving at their origin stops and move with the buses to reach their destinations. The system must know at which bus stop a passenger can get off and which bus the passenger can take. In addition, a passenger may take a transfer at some stop if the origin and destination are on different routes. In this case, the system needs to select a stop for the passenger to wait. These decision-making processes are achieved by solving for the shortest path between the passenger's origin and destination through methods such as the Dijkstra’s algorithm \citep[see, e.g.,][]{ahuja1988network}. Functions GetOn, GetOff, and NextStop shown in Algorithm~\ref{alg_func} use the shortest path and the current stop to determine if a passenger needs to get on a bus, get off a bus, or select a stop to wait/exit, respectively. 

\begin{algorithm}[ht!]
\footnotesize
  \caption{Main logic process in the discrete-event simulation.}
    \begin{algorithmic}
    \IF {Entity is Bus}
    \IF {Location is LA} 
    \STATE Compute WaitTime 
    \STATE Load currCap Passenger From WA In WaitTime If GetOn(Passenger, currStop) $=$ Route 
    \STATE Compute currCap 
    \STATE Move To UA 
    \ENDIF 
    \IF {Location is UA} 
    \STATE Unload currCap Passenger If GetOff(Passenger, currStop) is True 
    \STATE Compute currCap 
    \STATE Move To LA of the next stop on the Route 
    \ENDIF 
    \ENDIF 
    \IF {Entity is Passenger} 
    \IF {Location is UA} 
    \STATE stopID $\leftarrow$ NextStop(Passenger, currStop) 
    \STATE Exit If stopID $= -1$ 
    \STATE Move To WA of stopID If stopID $\geq 0$
    \ENDIF
    \IF {Location is WA} 
    \STATE Move To LA If Load Requested and GetOn(Passenger, currStop) $=$ Route 
    \ENDIF
    \IF {Location is LA} 
    \STATE Loaded On Bus 
    \ENDIF 
    \ENDIF
    \end{algorithmic}
  \label{alg_main}
\end{algorithm}

\begin{algorithm}[ht!]
\footnotesize
  \caption{Functions used in the simulation to decide individual passengers' actions.}
    \begin{algorithmic}
    \STATE \textbf{function} Dijkstra(origin, dest):
    \STATE Input: route graph, stopID origin, and stopID dest
    \STATE Require: stop dest is accessible to stop origin
    \STATE Output: the shortest path between the given  stops: origin and dest
    \STATE \textbf{end function}
    \STATE \textbf{function} GetOn(Passenger, currStop):
    \STATE Input: Passenger's attributes (origin and dest), stopID currStop that Passenger is currently at
    \STATE Require: currStop is on the shortest path; Passenger is at WA
    \STATE Output: Route that Passenger can take to move on the shortest path
    \STATE \textbf{end function}
    \STATE \textbf{function} GetOff(Passenger, currStop):
    \STATE Input: Passenger's attributes (origin and dest), stopID currStop that Passenger is currently at
    \STATE Require: currStop is on the shortest path; Passenger is on bus
    \STATE Output: True if currStop is the destination stop or a transfer stop
    \STATE \textbf{end function}
    \STATE \textbf{function} NextStop(Passenger, currStop):
    \STATE Input: Passenger's attributes (origin and dest), stopID currStop that Passenger is currently at
    \STATE Require: currStop is on the shortest path; Passenger is at UA
    \STATE Output: stopID of the next stop to wait;  $-1$ (Exit) if currStop is the destination stop; 
    \STATE \textbf{end function}
    \end{algorithmic}
  \label{alg_func}
\end{algorithm}

\normalsize
\subsection{Inputs and Parameter Settings}

We consider the six routes given by Model \eqref{ts-mip}, their frequencies and the number of buses in each route designed in the previous section. 
We present detailed stops on the six routes, their travel time and the bus stop time in Tables \ref{main}--\ref{north3} in the Appendix. 
The actual stop time of a bus at each stop in our simulation follows the following rules: It waits 1 minute if it needs to load/unload a passenger. It waits 2 minutes at the two hubs (in Central and North campuses) or at a stop having $> 10$ passengers who need to be loaded to or unloaded from the bus. These numbers are given by our collaborator and based on their experiences.

In the simulation, we focus on two 2-hour periods: 8am-10am and 12pm-2pm, which are considered  by our collaborator ``the busiest hours in either Fall or Winter semester at UM during pre-COVID years,'' and our goal is to make sure that the new system can meet demand volumes if they were similar to pre-COVID semesters, but only using 50\% or less of the total number of seats on each bus. In addition to metrics related to bus operations (e.g., bus capacity utilization and their operating time), we are also interested in measuring the passenger-related metrics, including their aggregated waiting time and the number of transfers needed for completing their trips.

The mean and variances of the exponentially distributed random arrivals capture the variability in the volumes of passengers at different stops, and they are based on real UM student activity data in 2019. (For details about the data sources we use and types of parameters estimated using those data, please see our earlier discussions in the ``Mathematical Models for Route Design'' section.) The 8am-10am and 12pm-2pm periods we test contain key threshold time such as class starting and ending time, and we sample random arrivals of passengers following an exponential distribution with mean computed from empirical data. On average, we have 1750 passengers arriving at all stops per peak hour (during 8am-10am or 12pm-2pm) in pre-COVID semesters, and asked by our collaborator, we test the aggregated total demand $D=2625$ as $1.5\times1750$ in our baseline case, where we also consider 50\% bus capacity $C = 40$ seats per bus. As benchmarks, we test two other cases where $C = 20, D = 1500$ and $C = 20, D = 2625$ requested by our collaborator. They both correspond to only 25\% bus capacity (i.e., 20 seats) are allowed to be used and the corresponding peak hour demand being reduced to 1500 passengers at all stops or staying the same. 

We use the percentages of Origin-Destination (O-D) pairs shown in Table~\ref{arrival_8} and Table~\ref{arrival_12} (in the Appendix) for generating passenger arrivals and their trip destinations for 8am-10am and 12pm-2pm, respectively. The O-D pair distributions are based on the analysis of number of residents in different student dorms, their majors, daily activities, and course schedules in 2019-2020. We use $88\%$ of the total demand $D$ in each case  multiplied by the O-D distributions in Tables \ref{arrival_8} and \ref{arrival_12} to compute the mean values of the exponentially distributed arrivals at each origin and their destination distributions for sampling passenger arrivals.
Additionally, 12\% of the total passengers will randomly, uniformly arrive at one of the 44 stops and get off at any of the other 43 stops with equal probability. 

We assume that passengers can make transfers at the same stops or at tuples of nearby stops immediately. Some pairs or triples of stops are considered to be within walking distance, so that passengers can make a transfer within 0.5 minute. In the Appendix, we present the list of bus stops that allow same-stop transfer in Table~\ref{same_stop} and the list of bus stops that allow nearby-stop transfer in Table~\ref{near_stop}.

We test the new bus system’s performance for the three aforementioned cases with different bus capacity $C$ and total demand $D$. We also conduct tests about the system's resilience when there exists a random bus break-down on any of the six routes. We run 40 replications for each case and use ``average utilization rate per stop-to-stop trip" $U_s$ and ``percentage stop-to-stop trip with at least 75\% utilization rate" $S_{0.75}$ to measure bus operational performance. A stop-to-stop trip is defined as the buses’ travel for one stop, i.e., from the previous stop to the next stop along a route. When a bus leaves a stop, we compute its utilization rate as
\begin{equation}
    U = \frac{\text{currCap}}{\text{maxCap}}, 
\end{equation}
and then calculate $U_s$ as the accumulated utilization rate divided by the total number of stops that buses have passed by. To obtain $S_{0.75}$, we count the percentage of stop-to-stop trips that have utilization rate no less than 75\%. 

For passenger-related metrics, we focus on the time each passenger spends in the system on average and how many transfers they take to complete their trips. We also measure the average wait time of each passenger spent at each bus stop. If a passenger needs to transfer to a different bus, the wait time is the sum of wait time at all the transfer stops. The time to walk to nearby stops is also included in the wait time.

\subsection{Result Analysis}
\label{sec:result}
In Table \ref{tab:main_results}, we show the results related to bus operations and passenger-related performance measures for the three cases. The maximum utilization rate of buses on any of the six routes is around 50\%-60\%, indicating that on average the number of passengers at each bus would be no more than 25. This satisfies the social-distancing requirement discussed in \citet{bluebus-aerosols} 
for public transit systems. In addition, no more than 54\% of buses load more than 75\% of its capacity in all three cases. Although increased $D$- and decreased $C$-values result in higher utilization of buses on each route, the differences are not significant. This justifies that the new system will not be overloaded even we require more strict social distancing and decrease the bus capacity from 50\% to 25\%. Overall, the average number of passengers who take one or multiple transfers is around 20\% and the average on-bus time of each passenger is between 12-13 minutes in all three test cases. However, the average waiting time increases to 14.46 minutes from 6.78 minutes if we reduce the capacity from 50\% to 25\% of the total seats on each bus. 

\begin{table}[ht!]
  \centering
  \caption{Results of bus- and passenger-related performance measures for the three test cases.}
    \resizebox{0.5\textwidth}{!}{%
    \begin{tabular}{lrrr}
    \hline
    \textbf{Test Case} & 1     & 2     & 3 \\
    \hline
   % \textbf{Settings} &       &       &  \\
    Bus Capacity ($C$) & 40    & 20    & 20 \\
    Total Demand ($D$) & 2625  & 1500  & 2625 \\
    \hline
    %\textbf{Bus-specified} &       &       &  \\
   \textbf{Bus Capacity Utilization}  $U_s$ &       &       &  \\
    All routes & 0.33  & 0.35  & 0.43 \\
    Campus Connector  & 0.22  & 0.24  & 0.29 \\
    Stadium-Diag Loop & 0.09  & 0.10   & 0.17 \\
    Oxford-Markley Loop & 0.28  & 0.31  & 0.47 \\
    Green Rd-NW5 Loop & 0.51  & 0.51  & 0.53 \\
    Bursley-Baits Loop & 0.54  & 0.56  & 0.61 \\
    Northwood Loop & 0.33  & 0.35  & 0.43 \\
    \hline
    %      &       &       &  \\
    \textbf{Percentage of Overloaded Buses}  $S_{0.75}$ &       &       &  \\
    All routes & 0.08  & 0.11  & 0.31 \\
    Campus Connector  & 0.47  & 0.49  & 0.54 \\
    Stadium-Diag Loop & 0     & 0     & 0.01 \\
    Oxford-Markley Loop & 0.38  & 0.38  & 0.39 \\
    Green Rd-NW5 Loop & 0.49  & 0.5   & 0.5 \\
    Bursley-Baits Loop & 0.15  & 0.17  & 0.19 \\
    Northwood Loop & 0.22  & 0.24  & 0.32 \\
    \hline
  %  \hline
   \textbf{Passenger-related Results} &       &       &  \\
    Avg.\ total travel time (min) & 19.29 & 20.68 & 26.59 \\
    Avg.\ wait time (min) & 6.78  & 8.42  & 14.46 \\
    Avg.\ on-bus time (min) & 12.51 & 12.26 & 12.13 \\
    Avg.\ number of transfers & 0.25  & 0.25  & 0.25 \\
    \% who make transfers & 20.10\% & 20.20\% & 20.00\% \\
    \hline
    \end{tabular}
    }
  \label{tab:main_results}
\end{table}

\begin{figure}
    \centering
    \includegraphics[width=0.9\textwidth]{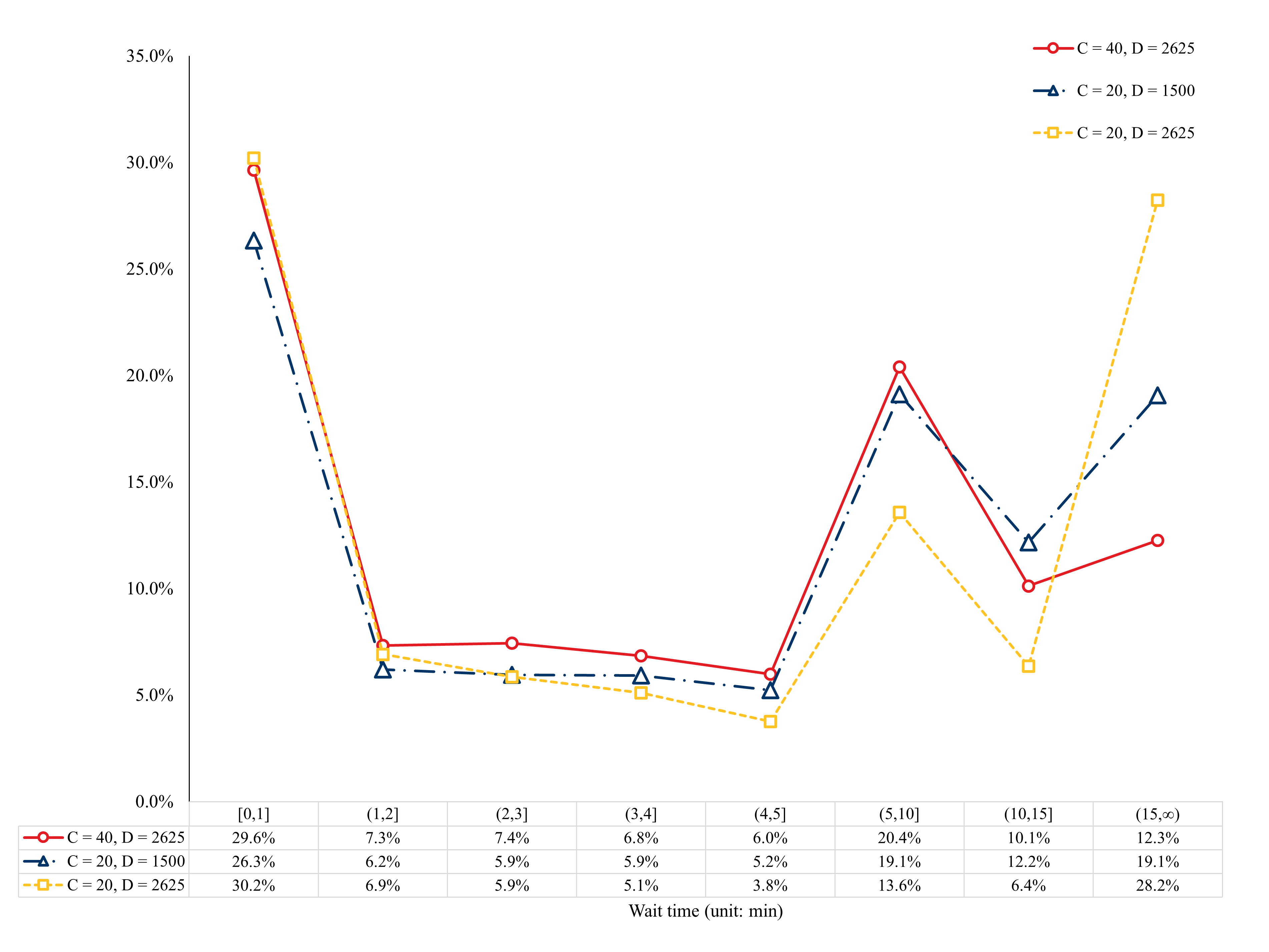}
    \caption{Percentages of passengers for different lengths of wait time in the three test cases.}
    \label{fig:wt_dist}
\end{figure}

In Figure~\ref{fig:wt_dist}, we further investigate passengers' wait time distribution in the three cases for  8am-10am and present the percentages of passengers waiting for different amounts of time. In all three cases, the average wait time is less than 15 minutes. More than half passengers wait less than 5 minutes. Only 1 passenger in 5 needs at least one transfer to complete a trip. In the third case when $C=20, D=2625$, there are significantly amounts of passengers waiting longer than 15 minutes (i.e., close to 30\%), which leads to significantly longer average waiting time in that case depicted in Table \ref{tab:main_results}. 

Next, we perform stress tests by randomly breaking down one bus (i.e., reducing the total number of buses by one) on one route in each case, while keeping other parameter values the same. Using the first case ($C = 40, D = 2625$) as the baseline, we present statistics of passenger-related performance measures in Table~\ref{tab:breakdown}. 

\begin{table}[htbp]
\centering
\caption{Results of one bus break-down on each route for the case $C=40, D=2625$.}

    \resizebox{\textwidth}{!}{%
\begin{tabular}{lccccccc}
\hline
Metric         & Baseline & Campus Connector & Stadium-Diag & Oxford-Markley & Green Rd-NW5 & Bursley-Baits & Northwood\\ \hline
Avg.\ total travel time (min)                   & 19.29    & 19.35            & 19.49             & 20.14               & 19.60             & 19.37              & 18.33          \\
Avg.\ wait time (min)          & 6.78     & 6.95             & 7.06              & 8.02                & 7.19              & 6.92               & 6.19           \\
%Wait $\geq$  5 min (\%) & 48.77\%  & 49.67\%          & 49.74\%           & 54.51\%             & 51.19\%           & 49.32\%            & 53.14\%        \\ \hline
Wait $\geq 5$ min (\%)   & 42.78\% & 43.57\% & 43.63\% & 47.81\% & 44.90\% & 43.26\% & 46.61\%\\
\hline
\end{tabular}
}
  \label{tab:breakdown}
\end{table} 

The performance results in Table \ref{tab:breakdown} are similar to the ones in Table \ref{tab:main_results} and Figure \ref{fig:wt_dist} for all three test cases, which implies that the system is insensitive to small changes of bus availability on any route and it is resilient to capacity reduction. 

\section{Post-Implementation Discussion and Conclusion}

The purpose of the simulation was to verify whether the designed new routes can possibly handle demand if they were as high as from pre-covid time (i.e., the data we used in the simulation is from year 2019-2020) and we report the corresponding results of bus utilization and users’ experiences (such as waiting time and number of transfers they take) to show the consequences of shortening all the routes. The new bus system was implemented at UM Ann Arbor campuses during Fall 2020 and Winter 2021 semesters, during which, individual passengers' riding time, waiting, and transfer time are reported comparable to the simulation results, while bus utilization rates are much lower because the estimated student demand is much higher than the actual demand during 2020-2021. The buses mainly served demand from medical school and we were able to ensure 50\% or lower bus capacity use to allow social distancing during peak hours. We also made sure that all passengers spent no more than 15 minutes on each bus. The hub-and-spoke design meets the requirements from UM policymakers during the most severe months of the COVID-19 pandemic (especially when COVID vaccines were not available). 

In Fall 2021, the university requires all student, faculty and staff being vaccinated and all need to be masked when indoor or on buses. The 50\% capacity limit, 15-minute single-trip limit, and social distancing requirement are lifted due to the new vaccination and mask policy. Moreover, due to driver shortage, the buses cannot run the same frequency designed in this paper, and we will return to normal bus routes used before COVID.

A hub-and-spoke transportation system relies on short, direct routes with fewer number of stops. The shorter routes can enable more frequent services for public transit systems, to reduce passenger wait time as well as the number of passengers on each vehicle. Through studying how particles exhaled from passengers travel in a vehicle under various conditions \citet{bluebus-aerosols} developed guidelines for operating campus buses more safely during the COVID-19 pandemic. Among these guidelines, each UM campus bus cannot hold more than 50\% of the total capacity (or even lower) and each passenger is desired to ride no more than 15 minutes in the same bus. In this work, we redesigned the UM campus bus system to (i) shorten routes that connect all UM Ann Arbor campuses and (ii) increase overall bus capacity utilization and frequency. We consolidated bus stops to reduce the number of stops. We estimated critical demand for essential bus stops and O-D pairs, built an integer programming model to generate initial routes, run simulation to evaluate the performance and resilience of the new design, and also improved it by identifying bottlenecks through stress-tests. As it is essential to enforce social distancing to control virus spread during the pandemic, our approach can be utilized to quickly redesign public transit systems in different scales, for temporary or permanent use.

~\\
{\bf Acknowledgements:}
The authors are grateful for the support from University of Michigan (UM) Logistics, Transportation and Parking Department, for supplying data of the UM campus bus system. The authors also thank the College of Engineering at UM for administrative support to the team and financial support to all students who are involved in the research during Summer 2020.



%\bibliographystyle{abbrvnat}
%\bibliographystyle{plainnat}
%\bibliographystyle{informs2014} % outcomment this and next line in Case 1
%\bibliography{bluebus} % if more than one, comma separated

\appendix

\section*{APPENDIX}
\section{Information Used in the Simulation}
\label{sec:appen}

Tables \ref{main}, \ref{central1}, \ref{central2}, \ref{north1}, \ref{north2} and \ref{north3} below show the bus stops on each new route, travel time from one stop to the next, and average bus stop time at each stop used in our simulation.  

%\small
\begin{table}[ht!]
  \centering
  \caption{Stops on route Campus Connector}
  \resizebox{0.6\textwidth}{!}{%
    \begin{tabular}{lcc}
    \hline
    Stop  & \multicolumn{1}{l}{Travel Time  to Next Stop (min)} & \multicolumn{1}{l}{Avg.\ Bus Stop Time (min)} \\    \hline
    Pierpont - Bonisteel & 1.7   & 2 \\
    Mitchell Field (Lot NC 78) & 2.9   & 1 \\
    Glen and Catherine Inbound & 2.2   & 1 \\
    Museum & 0.9   & 2 \\
    Power Center & 1.0   & 1 \\
    Glen and Catherine Outbound & 2.9   & 1 \\
    Mitchell Field Lot M75 & 2.8   & 1 \\
    Cooley - Inbound & 0.7   & 1 \\
    Pierpont - Bonisteel &  --     & -- \\
    \hline
    \end{tabular}
    }
  \label{main}
\end{table}

\begin{table}[ht!]
  \centering
  \caption{Stops on route Stadium-Diag Loop}
  \resizebox{0.6\textwidth}{!}{%
    \begin{tabular}{lcc}
    \hline
    Stop  & \multicolumn{1}{l}{Travel Time  to Next Stop (min)} & \multicolumn{1}{l}{Avg.\ Bus Stop Time (min)} \\
    \hline
    Oxford Housing & 1.7   & 1\\
    Stockwell & 3.0   & 1 \\
    Mott Inbound & 1.8   & 1 \\
    BSRB  & 1.5   & 1 \\
    Rackham & 1.5   & 1 \\
    CCTC - Chemistry & 1.7   & 2 \\
    East Quad & 1.2   & 2 \\
    Henderson House & 2.2   & 1 \\
    Oxford Housing &    --   &  -- \\
    \hline
    \end{tabular}%
    }
  \label{central1}%
\end{table}%

\begin{table}[ht!]
  \centering
  \caption{Stops on route Oxford-Markley Loop}
  \resizebox{0.6\textwidth}{!}{%
    \begin{tabular}{lcc}
    \hline
   Stop  & \multicolumn{1}{l}{Travel Time  to Next Stop (min)} & \multicolumn{1}{l}{Avg.\ Bus Stop Time (min)} \\
    \hline
    Crisler Center Lot SC-7 & 2.0   & 1 \\
    Kipke and Green & 1.8   & 1 \\
    IM Building Outbound & 2.3   & 2 \\
    Law Quad & 0.7   & 1 \\
    Michigan Union (NB State) & 2.0   & 2 \\
    Kraus & 1.3   & 1 \\
    CCTC - Chemistry & 2.4   & 2 \\
    East Quad & 2.5   & 2 \\
    Hill at Oakland & 3.3   & 1 \\
    New Inbound IM & 1.6   & 1 \\
    ICLE  & 5.2   & 1 \\
    Crisler Center Lot SC-7 &  --    &  -- \\
    \hline
    \end{tabular}%
    }
  \label{central2}%
\end{table}%

\begin{table}[ht!]
  \centering
  \caption{Stops on route Green Rd-NW5 Loop}
  \resizebox{0.6\textwidth}{!}{%
    \begin{tabular}{lcc}
    \hline
    Stop  & \multicolumn{1}{l}{Travel Time  to Next Stop (min)} & \multicolumn{1}{l}{Avg.\ Bus Stop Time (min)} \\    \hline
    Green Road Park and Ride & 3.6   & 2 \\
    Northwood 5 & 0.8   & 2 \\
    NCAC (on Hubbard) & 0.7   & 1 \\
    Hubbard and Hayward - Inbound & 1.1   & 1 \\
    FXB - Inbound & 1.3   & 1 \\
    LMBE  & 2.1   & 1 \\
    Art and Architecture & 2.3   & 2 \\
    FXB - Outbound & 1.4   & 1 \\
    Hubbard and Hayward - Outbound & 1.0   & 1 \\
    NCAC  & 1.9   & 1 \\
    Northwood 5 & 4.2   & 2 \\
    Green Road Park and Ride &   --    & -- \\
    \hline
    \end{tabular}%
    }
  \label{north1}%
\end{table}%

\begin{table}[ht!]
  \centering
  \caption{Stops on route Bursley-Baits Loop}
  \resizebox{0.6\textwidth}{!}{%
    \begin{tabular}{lcc}
    \hline
   Stop  & \multicolumn{1}{l}{Travel Time  to Next Stop (min)} & \multicolumn{1}{l}{Avg.\ Bus Stop Time (min)} \\    \hline
    Pierpont-Murfin & 2.3   & 2 \\
    Baits 2 & 0.8   & 2 \\
    Bursley & 1.3   & 2 \\
    Northwood 1 & 1.9   & 1 \\
    Hubbard/Hayward Lot (NC46) & 1.6   & 2 \\
    FXB Inbound & 2.6   & 1 \\
    Pierpont-Murfin &     --  &  -- \\
    \hline
    \end{tabular}%
    }
  \label{north2}%
\end{table}%

\begin{table}[ht!]
  \centering
  \caption{Stops on route Northwood Loop}
  \resizebox{0.6\textwidth}{!}{%
    \begin{tabular}{lcc}
    \hline
    Stop  & \multicolumn{1}{l}{Travel Time to Next Stop (min)} & \multicolumn{1}{l}{Avg.\ Bus Stop Time (min)} \\
    \hline
    Pierpont - Murfin & 1.3   & 2 \\
    Northwood 3 & 1.9   & 1 \\
    Northwood 2 & 1.7   & 2 \\
    Plymouth Road Crosswalk & 1.1   & 2 \\
    Northwood Community Center & 0.6   & 2 \\
    Hubbard/Hayward Lot (NC46) & 3.4   & 2 \\
    Pierpont - Murfin & --      & -- \\
    \hline
    \end{tabular}%
    }
  \label{north3}%
\end{table}%

Tables \ref{arrival_8} and \ref{arrival_12} below demonstrate the percentages of passenger arrivals at each stop and the distributions of their destinations used in the simulation, during 8am-10am and 12pm-2pm periods, respectively.  

\begin{table}[ht!]
  \centering
  \caption{Passenger arrivals during 8am-10am}
    \resizebox{0.6\textwidth}{!}{%
    \begin{tabular}{lrr}
    \hline
    Origin & \multicolumn{1}{l}{Percentage (\%)} & \multicolumn{1}{l}{Destination}\\
    \hline
    Plymouth Road Crosswalk & 7.08 & \multicolumn{1}{l}{Pierpont - Murfin} \\
          & 1.92 & \multicolumn{1}{l}{Museum} \\
    Northwood Community Center & 2.36 & \multicolumn{1}{l}{Pierpont - Murfin} \\
          & 0.64 & \multicolumn{1}{l}{Museum} \\
    Northwood 2 & 2.36 & \multicolumn{1}{l}{Pierpont - Murfin} \\
          & 0.64 & \multicolumn{1}{l}{Museum} \\
    Northwood 5 & 4.55 & \multicolumn{1}{l}{Art and Architecture} \\
          & 2.45 & \multicolumn{1}{l}{Museum} \\
    Michigan Union (NB State) & 14  & \multicolumn{1}{l}{Pierpont - Bonisteel} \\
    CCTC  & 19   & \multicolumn{1}{l}{Pierpont - Bonisteel} \\
    Baits 2 & 12  & \multicolumn{1}{l}{Museum} \\
    Bursley & 8  & \multicolumn{1}{l}{Museum} \\
    IM Building Outbound & 1.35 & \multicolumn{1}{l}{CCTC - Chemistry} \\
          & 1.5 & \multicolumn{1}{l}{Pierpont - Murfin} \\
    Green Road Park and Ride & 4.2 & \multicolumn{1}{l}{FXB Inbound} \\
          & 2.8 & \multicolumn{1}{l}{Hubbard and Hayward - Inbound} \\
    Random & 12  & \multicolumn{1}{l}{Random}\\
    \hline
    Total & 100     &  \\
    \hline
    \end{tabular}%
    }
  \label{arrival_8}%
\end{table}%

\begin{table}[ht!]
  \centering
  \caption{Passenger arrivals during 12pm-2pm}
    \resizebox{0.6\textwidth}{!}{%
    \begin{tabular}{lrr}
    \hline
    Origin & \multicolumn{1}{l}{Percentage (\%)} & \multicolumn{1}{l}{Destination}\\
    \hline
    Pierpont - Murfin & 7.08 & \multicolumn{1}{l}{Plymouth Road Crosswalk}\\
          & 2.36 & \multicolumn{1}{l}{Northwood Community Center} \\
          & 2.36 & \multicolumn{1}{l}{Northwood 2} \\
          & 1.5 & \multicolumn{1}{l}{IM Building} \\
    Pierpont - Bonisteel & 10   & \multicolumn{1}{l}{CCTC - Chemistry} \\
          & 14  & \multicolumn{1}{l}{Michigan Union (NB State)} \\
    Art and Architecture & 4.55 & \multicolumn{1}{l}{Northwood 5} \\
    CCTC - Chemistry & 1.35 & \multicolumn{1}{l}{New Inbound IM} \\
    Museum & 1.92 & \multicolumn{1}{l}{Plymouth Road Crosswalk} \\
          & 0.64 & \multicolumn{1}{l}{Northwood Community Center} \\
          & 0.64 & \multicolumn{1}{l}{Northwood 2} \\
          & 2.45 & \multicolumn{1}{l}{Northwood 5} \\
          & 12  & \multicolumn{1}{l}{Baits 2} \\
          & 8  & \multicolumn{1}{l}{Bursley} \\
    FXB Outbound & 4.2 & \multicolumn{1}{l}{Green Road Park and Ride} \\
    Hubbard and Hayward - Outbound & 2.8 & \multicolumn{1}{l}{Green Road Park and Ride} \\
    Random & 12  & \multicolumn{1}{l}{Random}\\
    \hline
    Total & 100     & \\
    \hline
    \end{tabular}
    }
  \label{arrival_12}
\end{table}

Tables \ref{same_stop} and \ref{near_stop} below show the bus stops passengers use for same-stop transfers or nearby transfers, and their corresponding routes. 

\begin{table}[ht!]
  \centering
  \caption{Stops where same-stop transfers are available}
    \resizebox{0.6\textwidth}{!}{%
    \begin{tabular}{lll}
    \hline
    Stop  & Route 1 & Route 2 \\
    \hline
    CCTC-Chemistry & Stadium-Diag Loop & Oxford-Markley Loop \\
    East Quad & Stadium-Diag Loop & Oxford-Markley Loop \\
    FXB Inbound & Bursley-Baits Loop & Green Rd-NW5 Loop \\
    Pierpont - Murfin & Northwood Loop & Bursley-Baits Loop \\
    Hubbard/Hayward Lot 46 & Northwood Loop & Bursley-Baits Loop \\
    \hline
    \end{tabular}
    }
  \label{same_stop}
\end{table}

\begin{table}[ht!]
  \centering
  \caption{Stops that allow nearby-stop transfers}
    \resizebox{0.7\textwidth}{!}{%
    \begin{tabular}{rll}
    \hline
    \multicolumn{1}{l}{Location} & Stops & Passing Routes \\
    \hline
    \multicolumn{1}{l}{CCTC} & Chemistry & Stadium-Diag Loop, Oxford-Markley Loop \\
          & Museum & Campus Connector \\
    \multicolumn{1}{l}{FXB Building} & Inbound & Green Rd-NW5 Loop \\
          & Outbound & Green Rd-NW5 Loop \\
    \multicolumn{1}{l}{NCAC} & Hubbard & Green Rd-NW5 Loop \\
          & South Outbound & Green Rd-NW5 Loop \\
    \multicolumn{1}{l}{Pierpont} & Bonisteel & Campus Connector \\
          & Murfin & Bursley-Baits Loop, Northwood Loop \\
          & Art \& Architecture & Green Rd-NW5 Loop \\
    \multicolumn{1}{l}{Admin. Service} & Kipke and Green & Stadium-Diag Loop  \\
          & ICLE  & Stadium-Diag Loop \\
    \multicolumn{1}{l}{Mitchell Field} & Lot NC - 78 & Campus Connector \\
          & Lot M - 75 & Campus Connector \\
    \multicolumn{1}{l}{Glen and Catherine} & Inbound & Campus Connector \\
          & Outbound & Campus Connector \\
    \multicolumn{1}{l}{IM Building} & New Inbound & Oxford-Markley Loop \\
          & Outbound & Oxford-Markley Loop \\
    \multicolumn{1}{l}{Hubbard/Hayward} & Lot NC - 46 & Bursley-Baits Loop, Northwood Loop \\
          & Inbound & Green Rd-NW5 Loop \\
          & Outbound & Green Rd-NW5 Loop \\
    \hline
    \end{tabular}
    }
  \label{near_stop}
\end{table}

\end{document}